\documentclass[12pt]{article}     
  \usepackage{graphics}
  \usepackage{flafter}

\begin{document}
\title{ Why Einstein (Had I been born in 1844!)?}
\author{Naresh Dadhich \footnote{email: nkd@iucaa.ernet.in}\\
IUCAA, Post Bag 4, Pune ~ 411~007, India}
\maketitle 
\begin{abstract}                      
In his monumental discoveries, the driving force for Einstein was, I believe, consistency 
of concept and principle rather than conflict with experiment. In this spirit, I would like 
to look at the journey from the classical to the relativistic world as a simple and direct 
exercise first in recognition of universal character of universal entities and then carrying 
out the universalization. By this process not only the relativistic world follows most 
naturally but I would like to conjecture that if Einstein were born in 1844 (or had I 
been born in 1844 and had followed this line of thought as I do now!) it would have in 
fact been predicted including existence of a wave with universal constant velocity. That would 
have indeed been not only the greatest but most amazing and remarkable feat of human thought.  

Beating further on the same track of principle and concept driven ideas, we 
ponder over to see beyond Einstein, and ask the questions: in how many dimensions does gravity live, how many basic forces are there in nature and what are the basic building blocks of space-time? 

\end{abstract}
 
\section{Introduction}

Let me begin by hypothesizing that it is tempting to place Einstein or oneself in the 
pre-Maxwell times and follow the natural line of thought which is in the Einsteinian 
spirit, and see what happens. Further in the present context, what type of questions does this line of thought give rise to? In this essay, I wish to follow the Einsteinian spirit of consistency of principle and concept, and let the rest follow naturally all by itself. 

We shall begin by defining universal entities and then identifying the primary examples of them. The rest of the story is built on seeking relation between the universal entities and universalization of concepts.

Let us begin at the very beginning by defining universality and the process of 
universalization. The natural definition for universal is that it is the same 
for all and shared by all. It could be an entity or concept like space, or could be a 
force like gravity. Since universal entity is the same and shared by all, this 
means all universal things must be related. That is, no two universal things can be 
independent. Any feature that distinguishes one universal thing from the other will mean that there exists some property which is not the same for all. This will violate the universal character. If there exist two universal entities or 
concepts, they must thus be related and the relation would have naturally to be universal. 

 \section{Primary Universal Entities} 

What are the most primary universal entities we know of? The natural and obvious answer 
is space and time. They are indeed the same and shared by all things that exist in 
nature. The two questions arise: since both are universal, hence one, they must be on the 
same footing and second, there must exist a universal relation between them. 

We know that the distance between two points depends upon the path an observer 
takes in going from one point to the other point in space. It is a common experience that 
kilometer reading in a taxi is path dependent and that is why we are quite watchful 
that the driver takes the shortest and not the circuitous path. Thus spatial distance is 
path dependent 
and so must be the time interval between any two events. This is what would be required 
to bring the two universal entities, space and time, on the same footing. This is however 
not so in the familiar Newtonian world. If we are to 
adhere steadfastly to our concept of universality, we are forced to seek new framework. 

Secondly, as argued above since both space and time are universal, there must exist a universal relation between them. The most natural relation between them is through velocity (which is 
formally defined by,  Velocity = Space/Time). Therefore a universal velocity is required which is 
the same for all observers irrespective of their relative motion. In the Newtonian 
mechanics, 
velocities are added by the law $w = u + v$. For instance, let two cars have velocity $u$ and 
$v$ relative to you on the ground, the relative velocity between the cars will be $w = u + 
v$. Existence of a universal velocity (which is the same for all and hence if one of $u$ and 
$v$ is the universal velocity, $c$, then $w$ must also be $c$) cannot be compatible with the 
Newtonian law of addition of velocity and consequently with the Newtonian mechanics. 
It therefore asks for a new mechanics. 

Thus we need a new mechanics simply by appealing to the universal character of space and 
time. One of the natural consequences of existence of the universal velocity is that 
like spatial distance between two points, time interval between any two events will now become path 
dependent as required for space and time to be on the same footing. Existence
of the universal velocity thus addresses both the above questions.

\section{ Universal Velocity} 

What should characterize the universal velocity, $c$ ? 

 $\bullet$ An object moving with $c$ should always be moving, never at rest relative to any observer.

 $\bullet$ It should be a limiting velocity for all observers, no observer can attain $c$. 

 $\bullet$ Since an object moving with $c$ can never be at rest, it cannot have non-zero (rest) mass. Its mass is zero. 

 $\bullet$ Existence of universal velocity means existence of zero mass particle. 

What physical phenomenon could provide universal velocity? To address this question, let us get at the root of the phenomenon of motion. It is of 
two kinds, particle-like and wave-like. A particle can be given arbitrary velocity, could 
be held at rest while a wave's motion is entirely determined by the elastic properties of 
the medium in which it propagates, it is always moving and can never be held at rest. A 
wave's velocity could thus be changed only either by changing or moving the medium. 
That means a wave propagating in a universal medium, which could neither be changed nor 
moved, will have universal velocity, the same for all observers. 

What is the universal medium? Obviously, space free of all matter called vacuum. Hence 
the universal 
velocity we are seeking could then only be provided by a wave propagating in free space 
(vacuum). Simply on the force of consistency of the concept, we thus make a profound 
prediction that there must exist a wave propagation in free space and it would have a universally constant velocity relative to all observers. 

\section{Had I Been Born in 1844!} 

Had I been born in 1844, it could have been quite possible to follow this 
train of thought. Then in around 1870, I could have, as a Young man of 26, as old as Einstein was 1905, made the above profound prediction 
that there must exist a wave propagating in free space with a universal constant velocity. That 
would have been a good 5 years before the Maxwell's electrodynamics. It would have been simply remarkable that a prediction made on the force of pure thought and concept comes actually true 5 years later in the form of electromagnetic wave of the Maxwell's electromagnetic theory. Its existence is then experimentally established by Hertz. Light turns out to be the most familiar example this 
new wave and thus its velocity becomes a universal constant. 

A great conjecture - I shudder to imagine what would have happened if the events did 
happen as projected. Forget me, let us place Einstein as a young man in 1870, what he was 
in 1905, he could have very well argued as we did above and would have come up with the 
profound prediction before Maxwell's theory. Had that happened, it would have been a great display of 
pinnacle of human thought and Einstein as its purest human manifestation! 

However things don't happen like this but it is insightful to wonder and probe the 
potential of sheer thought and analysis at its most pristine and sublime. 

\section{New Mechanics} 

A new mechanics could have thus been predicted by the general principle of universality 
that all universal entities be on the same footing and be related by a universal 
relation. This leads to existence of a universal constant velocity and the incorporation 
of this fact gives rise to a new mechanics which goes by the name, Einstein's Special 
Relativity. This year is being celebrated as the Year of Physics in due recognition of 
the fact that it marks the 100th anniversary of this monumental discovery 
(Simultaneously with this, he 
also made two other important discoveries of the photoelectric effect, which was just 
good enough to fetch him the Nobel Prize, and of the Brownian motion. But in the 
absence of this, they won't have merited this universal acclaim). 

The message of the new mechanics is that space and time are synthesized into one 
4-dimensional entity, space-time. Further it leads to the synthesis of mass and energy 
through the famous equation, $ E = Mc^2$, which has become synonymous with Einstein and much 
to his anguish and horror also with atom bomb. Since velocity of light, which is 
an electromagnetic wave, is universal - the same for all observers, it is the limiting 
speed of communication between spatially 
separated events in space. It therefore defines the velocity of causation. Events then 
get classified into two classes, those which can be causally connected and those which 
cannot be. For instance, light takes $8$ minutes to travel from the sun to the earth. 
The event 
which occurred at this moment on the sun can affect (be causally connected) events on the 
earth only after $8$ minutes have elapsed. That is, the events occurring on the earth between now and 
until $8$ minutes can have no connection (cannot be influenced by) with the this event 
on the sun. Thus we have finite velocity of physical communication - causal relation 
(causality), which is provided by the universal character of velocity of light. 

The rest of the Einsteinian mechanics is straightforward.

Another way of motivating Special Relativity could have been universalization of 
mechanics for all particles, massive as well as massless \cite{n1,n2}. The existence of massless 
particle would have as argued above provided the universal constant velocity. Once that 
happens, Special Relativity is inevitable. 

\section{New Gravity} 

The next natural question is to universalize gravity. That is, it acts on everything 
including massive as well as massless particles. A particle of light, called photon is 
an example of zero mass particle. The massless particle, photon propagates  always with 
the universal constant velocity which cannot change. Photon's velocity in vacuum is 
constant - a universal constant and hence it can never change. On the other hand, action 
of a force on anything is measured only through change in its velocity (motion). Velocity could however remain constant for circular motion but light does not always move in circle instead in the Newtonian theory it always moves in a straight line. 

We have thus landed with a serious contradiction in principle. Since gravity is 
universal, it must act on massless photon as well, yet its velocity must not change. How 
is it possible? In the classical Newtonian framework, it is impossible to reconcile with 
these two opposing properties. What should we do? We have no other go except to expand the framework \cite{n3}. How do we do that? 
   
When one is confronted with such a question of concept, it is only the robust common 
sense that can show the way. Let me take an uneducated peasant as personification of 
uninhibited common sense. I ask him this question. After some ponder, he says that he 
can't much appreciate action of gravity on light, and asks back what would you want  
light to do in actuality to feel gravity? I reply that if light is grazing past a massive 
body, it should bend toward it as every other thing does. He breathes deep and hard,
scratches at his beard and then asks me to follow him, and takes me to the river which 
flows in the backyard of the village. He picks up a piece of log and throws that into the 
river. It starts floating freely with the flow of the river. 

Then we walk along the river which suddenly bends, and so does the log. He smiles and 
inquires, did any force act on the log? No, I say, it is freely floating. But even then 
it bent with the river? He then triumphantly inquires, haven't you got the answer to your 
question? Dumb as I am, I say, No. He asks me where does your light float? In space, I 
say. Then what is the problem, he says, bend the space. Then it illuminates in me 
that why can't gravity bend/curve space around the massive body, and all particles 
including massless photons propagate/float freely in the curved space. In fact curved 
space-time, as space and time have already been synthesized into one in Special Relativity.

Isn't this simply astounding? 

What have we arrived at? A new theory of gravity where it can honestly be described by 
no other means than the space-time curvature itself. Gravity no longer remains a force 
but gets synthesized with structure of space-time and it simply becomes a property of the  
space-time geometry. Its dynamics, in particular the inverse square law of gravity,  
should now follow from the space-time curvature. That it does, all by itself. 
As Riemann curvature satisfies the Bianchi differential identities which on contraction yields the second rank symmetric tensor with vanishing divergence. This leads to the Einstein equation incorporating the Newtonian inverse square law. Nobody has to prescribe a law of gravity. We have thus 
discovered a new gravity - Einsteinian gravity, called by the name General Relativity. 

Gravity distinguishes itself from all other forces by the remarkable feature of impregnating space-time itself by its own dynamics. It then ceases to be an external force. 
Gravitational field is fully described by the curvature of space-time geometry. Motion under 
gravity will now be geodesic (straight line) motion relative to curved geometry of 
space-time. The geodetic motion should naturally include the Newtonian inverse square 
attractive pull, and in addition it should also have the effect of curvature of space. 
Light cannot feel the former but only feels the latter. Stronger is the gravity, stronger 
would be the curvature it produces. As we make the field stronger by making larger and 
larger mass confined to smaller and smaller region, it is conceivable that space gets so 
curved that light cannot propagate out but its orbit closes on itself around the massive 
body. When that happens a black hole is defined from which nothing can propagate out. 
Things can only fall into but nothing can come out - it defines a one-way membrane. This 
is the most remarkable and distinguishing prediction of the Einsteinian gravity. 

Let me make one point clear that despite all General Relativity literature being full of 
the phrase 'bending of light' due to gravity, it is as wrong as saying that the sun 
goes around the earth. What bends is space and light freely propagates in it. It was 
bending of space and not of light that was for the first time measured by Eddington in 
the 1919 total 
solar eclipse. We do however measure space bending by measuring deflection of light from 
the straight path. We do in fact see the sun going around the earth, yet we argue that it 
is apparent because we are sitting on the earth. Similarly, light can only freely 
propagate in space and can't bend, and hence its deflection is the measure of space bending.    
    
It is the universality of gravity which has first demonstrated the inadequacy of the 
Newtonian framework, and has then indicated the expansion of the framework by way of 
curving space-time such that the contradiction is resolved. That has led to the new Einsteinian gravity.
 
\section{In how many dimensions does gravity live?}

Everything about a universal force has to be determined all by itself - self determined. One has no freedom to prescribe anything. We have seen that it is the universality property which leads to space-time curvature for description of the universal force, and then of its own determined the equation of motion for 
the force - gravity. Next natural question that can arise for the universal force is, in how many dimensions should it live? The minimum dimensions required to define the curvature tensor is $2$, but then it is well-known that full dynamics of gravity cannot be realized in $2$ and $3$ dimensions. We thus come to the usual $4$ dimensions in which the matter fields sit in $3$ space dimensions. Why should matter remain confined to $3$-space, because its dynamics is fully 
realized in ($3+1$)-space-time. Hence there is no compelling physical reason to lift it to higher dimension. Should gravity then too remain confined to $(3+1)$-space-time? 

Gravity is a self-interactive force, and self interaction can only be computed iteratively. Since the space-time metric is the analogue of gravitational potential, first iteration of self interaction should involve square of its first derivative. This is however automatically included in the Einstein equation which follows from the Riemann curvature involving second derivative and square of first derivative. The question is, why should we stop at the first iteration, we should also go to next iterations? We have to get everything from the curvature tensor, should we square it? That will give higher powers of first derivative but will square second derivative as well. That is no good because the highest order of derivative must occur linearly in the equation to admit unique solution. If that is not the case, we will have more than one equation and hence more than one solution. This property is called quasi-linearity. We should thus seek the most general action for gravity constructed from the Riemann curvature and its contractions which leads to the second order quasi-linear differential equation.

Can we have second order iteration of self interaction and yet retain the quasi-linear character of the equation? The answer is yes. There exists a specific combination of square of Riemann, Ricci and Ricci scalar, called the Gauss-Bonnet combination which precisely does this. We should hence include the Gauss-Bonnet term which makes non-zero contribution only in dimension higher than $4$. This means physical realization of the second iteration of self interaction of 
gravity demands that it propagate in the extra $5$th dimension. It cannot remain fully confined to the $3$-space and its self-interaction dynamics takes it to leak into extra space dimension. This is an important conclusion which follows purely from classical consideration \cite{n4,n5}. It is remarkable that the differential geometry does provide a non-linear (in Riemann) action yet yielding the quasi-linear equation. 

Another motivation for extra dimension for gravity comes from the general principle of charge neutrality. A classical force should always be overall charge neutral like electromagnetic force. All bodies down to an atom are overall 
electric charge neutral. For gravity matter in any form defines charge which has positive energy and hence positive polarity. Gravitational charge is therefore unipolar. How do we now attain charge neutrality? The only option is to make field have charge of opposite polarity - negative. That is why universal force has to be self-interactive and always attractive! This is the most direct and 
simplest way to see why gravity is always attractive. Unlike electric charge, the negative polarity charge is non-localized and distributed with the field all over the space. If you integrate all over the space, negative charge of the field will fully balance the positive charge of matter field. However in the local neighbourhood around the matter distribution, there will be charge imbalance, over-dominance of positive charge, and hence the field has to propagate off the $3$-space locally. However it should not be able to go deeper in the extra dimension because as it propagates larger and larger region of negative charge
(field) will get included and hence its strength will diminish exponentially \cite{n3}. This means gravity essentially remains confined with massless mode having ground state on the $3$-space while propagation into the extra dimension 
is effectively through massive modes. Interestingly this is precisely the picture presented by the Randall - Sundrum brane world gravity \cite{rs}. Note that ours is purely a classical argument which does not take a priori existence of 
extra dimension but it is solely dictated by the dynamics of gravity. This is 
a very important difference. 

The curvature of space-time is not purely a geometric entity but it is the carrier of gravitational dynamics. So long as curved space-time is not isometrically embedded in higher dimensional flat space-time, it means that curvature does transmit non-trivial gravitational dynamics into the extra dimension and it cannot be flat. In general, a $4$-dimensional curved space-time is not isometrically embeddable in $5$-dimensional flat space-time unless it is conformally flat. Hence gravity cannot remain fully confined to $4$-space-time. This is purely a geometric argument for higher dimension. On the other hand, it has been shown that an arbitrary $4$-dimensional space-time can always be embedded in $5$-dimensional Einstein space (de Sitter or anti de Sitter) \cite{dr}. This again shows that once we hit conformally flat higher dimensional space-time, the iteration chain should terminate.  

Does this iteration chain stop or it goes on indefinitely to higher and higher dimensions? It naturally stops at the second iteration. In higher dimensional bulk space-time, only gravity propagates while matter fields remain confined to the $3$-space and hence the space-time is isotropic and homogeneous. It is then the maximally symmetric space-time of constant curvature which automatically solves the equation with the Gauss-Bonnet term and the Gauss-Bonnet parameter defining the constant $\Lambda$ in the bulk. It will be an Einstein space which is conformally flat with vanishing Weyl curvature. That means there exists no more any free gravity to propagate any further in higher dimension. Since Weyl curvature is zero, it can be isometrically embedded in higher $6$-dimensional flat space-time. The iteration chain thus stops at $5$-dimension. The important point to realize is that it is the property of gravity which naturally leads to higher dimension.

\section{How many basic forces should there be in Nature?} 

 Gravity is the unique universal force which is characterized by the following two properties: 

(i) Universal linkage, interaction with all particles massive as well as massless 

(ii) Long range, present everywhere.

It is shared by all that exist physically and hence could be taken as the mother force. All other forces should arise out of it by relaxing these two properties \cite{n1,n3}. How many different possibilities exist to give other forces in nature? 

There are naturally three possible options: one, relax (i) and retain (ii), two, retain (i) and relax (ii) and lastly relax both (i) and (ii). Including the mother, there could therefore exist only four basic forces in nature. This is a very simple unified view of all the four forces. Let us next see, do these possibilities conform with the forces we know of? 

In general long range forces will be classical while short range forces will be quantum and nucleus bound. There would exist two of them in each category. The mother force - gravity is a long range, classical tensor field. As we have seen, it is described by the curvature of space-time. The other long range classical force will arise on relaxing the property (i) of linkage to all. That means, this force will link to particles having a specific new parameter - charge. There should exist a new parameter other than mass/energy which will characterize this force. Again appealing to the principle of overall charge neutrality, the new charge will have to be bipolar (unipolar field could be only one for it has to be described by space-time geometry). This suggests that it will have to be a vector gauge field. It is a long range force and hence it should obey charge conservation which will lead to the usual inverse square law. All this clearly identifies the force to be Maxwell's electromagnetic with electric charge being the distinguishing charge parameter. Is this, like gravity, unique? Its dynamics is fully determined and hence there is no sensible reason for any other force to exist having the same dynamics as the electromagnetic field. Yes, it is indeed like gravity unique. 

Long range is characterized by massless and chargeless propagator so that it can freely propagate everywhere with the universal constant velocity. Short range on the other hand will be characterized by either propagator being massive and/or charged, or by coupling being running rather than constant. Short range forces will be quantum and will remain confined inside some region like the nucleus.

Now if we relax the property (ii) and retain (i);i.e. the force is short range but has universal linkage to all (massive) particles that can remain confined to short range - nucleus. The propagator is massive and/or charged and it negotiates interaction between massive electrically charged as well as neutral particles. It could be the weak force which interacts with all massive particles including neutrinos \cite{das}. Since this force interacts only with massive particles, it is predicted that neutrino must have non-zero mass. It is a kind of complementary ((i) but not (ii)) to the electric force ((ii) but not (i)). In an 
appropriate space there should exist a duality relation between them and the  electroweak unification is perhaps indicative of that. 

Lastly, let us relax both the properties, neither linkage to all (i) nor long range (ii). It is complementary to gravity which respects both the properties. It could be the strong force through which the smallest building blocks of matter, quarks interact with each-other \cite{das}. For the strong force, though propagator is massless but coupling is running which increases with distance and  vanishes as distance goes to zero. This feature is known as the asymptotic freedom. This property is dual of the asymptotic freedom of long range forces which become free at infinity.

Clearly, unlike the long range forces, it is not possible to establish uniqueness of the short range nuclear forces. If there exists a new force, this is the right place to look for it. Until we can establish their uniqueness, the question will remain open. The other question which this way of analysis prompts is that there should, in some appropriate space, exist duality based on the complementary character of the forces. That is between the electromagnetic and the weak force, and between the gravity and the strong force. They have complementary features. There are some suggestive indications. In the electroweak unification which is though not complete, it is significant that the two symmetry groups can be combined together as  product $SU(2)XU(1)$. On the other hand, there exists in the string theory, now the celebrated 
 $AdS/CFT$ (Anti de Sitter - conformal field theory) correspondence \cite{mal,witt}. At the boundary of the AdS space-time there lives the conformal field theory of matter - quantum chromodynamics. So we have AdS as gravity and CFT as the strong force dynamics of matter. This shows that there is a deeper connection between gravity and strong force. AdS/CFT correspondence is in fact reflection of the duality between the gravity (space-time) and the strong force (matter).

We would thus strongly argue for probing duality relations indicated by the complementary features of the forces. Hopefully, it may lead to some new insights. If nothing else, there emerges a unified picture of the four forces in a very simple manner. Unification of all the forces is the driving theme for the string theory which is also expected to give a quantum theory of gravity \cite{john}. String theory begins with the two basic universal principles, the special relativity and the quantum principle, and the rest follows from an attempt to construct a consistent theory of matter. First, it takes off to $26$ dimensions and then renormalizability of the theory brings it down to $10$ or $11$ dimensions. The pertinent question now is to come down to the $4$-dimensional space-time we live in. Unfortunately there is no unique way to do that. General Relativity appears along with plethora of other scalar fields in $4$ dimensions as an effective theory in the low energy limit.

\section{Seeing Beyond Einstein}  

The other great discovery, which signaled the turn of the 20th century, was that the 
micro structure of matter has quantum character - it is made of discrete tiny pieces. 
That is, matter is not continuous all the way down. As we make pieces smaller, the 
process of observing (probe) them will require smaller and smaller energy. Then there 
would occur the limiting situation where the energy of the object and the probe become 
comparable. It is again not reconcilable with the classical Newtonian 
framework and we need a new mechanics of Quantum Mechanics. 

This situation is characterized by the property that the process of observation disturbs 
the object non-ignorably. That is, a certain amount of uncertainty in observation becomes 
inherent which has to be incorporated in mechanics. That is, physical parameters will have to have probabilistic meaning and interpretation.  This realization is the key to quantum mechanics. The simplest manifestation of 
it is offered by the wave motion. On the one hand, it is characterized by the 
$4$-wavevector and on the other it should, like any other thing, carry energy and momentum 
which is given by the $4$-momentum vector. It stands to reason that the two $4$-vectors characterizing the 
same wave motion should be related, and this relation should be universal for it should 
be true for any wave motion. This leads to the basic relation between energy and 
frequency, and $3$-momentum and $3$-wave-vector, for which we have to introduce a new 
universal constant, called the Planck's constant. This constant is also the measure of 
the inherent uncertainty in measurement. It is obvious that it is impossible to localize a wave to 
determine its position precisely without making its momentum uncertain. Both position and momentum cannot simultaneously be determined to arbitrary accuracy. Accuracy of the one is only attainable at the cost of the other. This is the basic quantum principle, known as the uncertainty principle. The important message of quantum theory is that whenever object and probe 
become energetically comparable, motion tends to be wave-like. It could truly be described 
by quantum mechanics. At the micro scale, quantum mechanics thus becomes inevitable.    

With the discovery of electromagnetic wave and light its most common visible example, it 
is a pertinent and valid question to ask how does it propagate in vacuum? Should vacuum 
be completely physically inert or should it have some physical properties? This is a 
very contentious issue that brings forth the memory of infamous aether. Howsoever 
complex and involved the question be, we have to address it. 

Let us get our facts right. That is, what have been observationally established? One, 
an electromagnetic wave propagates in vacuum and second, one can't measure any 
motion relative to vacuum,i.e. it cannot act as a reference frame. It is the latter property 
which has been taken as evidence for untenability of the existence of aether. The former would naturally ask for 
oscillation of some basic constituents of the medium, vacuum to make wave propagate in 
it. Thereby it will, similar to matter, demand some kind of micro structure - basic 
building blocks of space. If that doesn't happen, nothing can propagate in it, nor can it 
bend as required by the Einsteinian gravity. Not only matter has quantum micro structure, 
space will also have to have some kind of structure. This is also required for 
physical realization of the quantum fluctuations of vacuum. If it has no micro structure, what will 
fluctuate? That means, even space cannot retain its continuum character all the way down.

Does this property necessarily have to conflict with the other, that it cannot act as a 
reference frame? No, because space is universal and hence is everywhere, it 
can't have anything of its own that could have a distinguishing character.  
Hence it can't define a reference frame for any motion. This is similar to a situation  that in middle of the ocean, without introducing a flag-post external to the ocean it is 
impossible to define a reference point. Ocean however does have micro structure yet it can't define a reference frame. Similarly, space itself can't define a reference frame but 
that can't stop it from having physical micro structure.  It is again the universality property  
of space which makes it reference-inert, even though it has micro structure. Hence there 
is no conflict between the two properties.

Quantum character is universal and hence space must also like matter acquire quantum 
behaviour at micro level. This view is further reinforced by the Einsteinian gravity in 
which space-time acquires physical property that it can bend like any other material 
medium. Recall the famous Wheelerian pronouncement, "Matter tells space how to curve 
and space tells matter how to move." Since space and matter respond to each-other 
physically, they should at a deeper level share the same property, or rather be on the 
same footing. 

Again the principle of universality demands that space-time should have quantum 
structure at micro level. The prime question then is, what are the basic building 
blocks of space-time? Nobody knows. That is indeed the most challenging question of the 
day. To address this question, we will have to find quantum theory of space-time itself. 
That is what is being done in the canonical approach to quantum gravity, called the loop quantum gravity \cite{lqg}. The Einsteinian gravity is described by the curvature of space-time, which is continuum but at the micro level it can no longer retain its continuum character. It has to turn into discrete quanta and hence we need a new theory of gravity which takes us beyond Einstein - a quantum theory of space-time and gravity.

On the other hand, the quantum principle (Uncertainty Principle) is universal and hence it must in some way be related with the most primary entity, space-time. It should really be deduced as a property of space-time. Until that happens, quantum theory will, in principle and concept, remain incomplete. The completion of quantum theory is also therefore asking for a new theory. That will perhaps come about by realizing the quantum character of space-time itself. That is to formulate quantum mechanics in a quantum space-time.

Doesn't it sound crazy and formidable? That it is. That is why the question has remained open for almost a century which is in good measure indicative of its complexity both in concept and technique.
It is interesting to note that both gravity and quantum theory are pointing at quantum character of space-time and its incorporation is thus imperative for new theory of gravity and quantum mechanics. That is the key and the most challenging question of the day. The two most prominent approaches addressing this fundamental question are the particle theory based string theory and the general relativity based canonical loop quantum gravity. There is however yet a long way to go before a complete theory of quantum space-time and gravity emerges.

\section{Outlook}   

The main theme of the essay is to demonstrate the power and sheer simplicity of pure  
thought 
and robust common sense. It is a tantalizing hypothesis that had a young person in 1860s 
wondered about things and argued as we have done above, it would have been quite possible 
to predict existence of a wave propagating with universal constant velocity in vacuum, 
and consequently the new mechanics of Special Relativity as well as new gravity 
of General Relativity. All this could have happened without any experiment challenging 
the existing theory. This was precisely what had happened for Einstein's discovery of 
General Relativity which was purely principle and thought driven. We are taking a one 
more step forward in that direction. Conceptually, all that what was required was available 
not only in 1860s but right from Newton's time. 

Why do I then single out 1860s? Because before that the prediction would have been too 
much ahead of its time, and hence would have had a still birth. Take the example of the 
greek philosopher, Aristarchus who is believed to have proposed that the earth goes 
around the sun and not the other way round as early as in 2nd century BC. This discovery couldn't 
survive for want of proper intellectual base and understanding which came into existence 
only in 15th 
century through the observation of planetary orbits. Then the time was ripe for Copernicus 
to make the monumental discovery. It is in this context that 1860s attain significance. 
With the Maxwell's theory of electromagnetism soon coming, the predicted wave 
would have been identified with the electromagnetic wave to be observed experimentally 
by Hertz. The stage was therefore well set for the profound prediction and discovery. 

It is interesting to conjecture and wonder, but the hard fact of life is that 
scientific discoveries are seldom driven, with the honorable and unique exception of 
General Relativity, by pure thought. They are essentially driven by contradiction 
between theory and observation, and the latter also depends upon the available technology 
for instruments. Truly, they are the products of the times - the prevailing scientific 
thought process. However had it happened as we envision, it would have been an amazing feat.

At any rate, it is a wonderful and straightforward way of looking at things 
based purely on simple logic and robust common sense. It is indeed insightful to wonder and ponder over the fundamental questions, such as the number of dimensions of space-time and number of basic forces in nature. In a natural way, one can argue that gravity requires minimum $4$-dimensions for full realization of its dynamics, and it also leaks into the extra dimension but cannot penetrate deep enough. Thus $4$ dimensions are necessary but not sufficient for gravity. Further, it is the unique universal force characterized by the properties; linkage to all and long range. By peeling off these properties, it is remarkable to have a unified view of all the four basic forces in nature and it also indicates why there are only four of them? The complementary features of the forces strongly 
suggest duality relation between gravity and the strong force, and electromagnetic and the weak force. This is one of the neat predictions of this way of analysis. Lastly it is important to realize that it is the completion of both quantum theory as well as general relativity that ask for a new quantum theory of space-time. The really challenging task is to do both gravity and quantum theory not in continuum but in quantum space-time.

\end{document}